\documentclass[aps,amsmath,amssymb,nofootinbib,showpacs,reprint]{revtex4-1}

\usepackage{amsmath}
\usepackage{amsfonts}
\usepackage{amssymb}
\usepackage{bbm}
\usepackage{slashed}
\usepackage{mathrsfs}
\usepackage[english]{babel}
\usepackage{microtype}
\usepackage{hyperref}
\usepackage{nicefrac}
\usepackage{graphicx}
\usepackage[babel]{csquotes}
\usepackage{color}
\usepackage{subcaption}

\DeclareMathAlphabet{\boldmathe}{T1}{cmr}{bx}{it}
\newcommand{\mbf}[1]{\boldmathe{#1}}

\newcommand{\N}{{\mathbb{N}}}
\newcommand{\Z}{{\mathbb{Z}}}

\newcommand{\ii}{\mathrm{i}}
\newcommand{\D}{\mathrm{d}}

\newcommand{\Nt}{N_\text{t}}
\newcommand{\Ns}{N_\text{s}}

\newcommand{\Nf}{N_{\text{f}}}

\def\vk{\mbf{k}}
\def\vn{\mbf{n}}

\newcommand{\fref}[1]{Fig.~\ref{#1}}

\newcommand{\gtapprox}{\raisebox{-0.5ex}{$\,\stackrel{>}{\scriptstyle\sim}\,$}}
\newcommand{\ltapprox}{\raisebox{-0.5ex}{$\,\stackrel{<}{\scriptstyle\sim}\,$}}
\begin{document}
\title{Baryons in the Gross-Neveu model in 1+1 dimensions at finite number of flavors}

\newcommand{\FSU}{Theoretisch-Physikalisches Institut, Friedrich-Schiller-Universit{\"a}t Jena, 07743 Jena, Germany}
\newcommand{\GU}{Institut f\"ur Theoretische Physik, Goethe-Universit\"at Frankfurt, Max-von-Laue-Stra{\ss}e 1, 60438 Frankfurt am Main, Germany}
\newcommand{\HFHF}{Helmholtz Research Academy Hesse for FAIR, Campus Riedberg, Max-von-Laue-Stra{\ss}e 12, 60438 Frankfurt am Main, Germany}

\author{Julian J.\ Lenz}
\email{julian.johannes.lenz@uni-jena.de}
\affiliation{\FSU}

\author{Laurin Pannullo}
\email{pannullo@itp.uni-frankfurt.de}
\affiliation{\GU}

\author{Marc Wagner}
\email{mwagner@itp.uni-frankfurt.de}
\affiliation{\GU \\ \HFHF}

\author{Bj\"orn H.\ Wellegehausen}
\email{bjoern.wellegehausen@uni-jena.de}
\affiliation{\FSU}

\author{Andreas Wipf}
\email{wipf@tpi.uni-jena.de}
\affiliation{\FSU}

\begin{abstract}
In a recent work \cite{Lenz:2020bxk} we studied the phase structure of the
Gross-Neveu (GN) model in $1+1$ dimensions at finite number of fermion flavors $\Nf = 2, 8, 16$, finite temperature and finite chemical potential using lattice field theory. Most importantly, we found an inhomogeneous phase at low temperature and large chemical potential, quite similar to the analytically solvable $\Nf \rightarrow \infty$ limit. In the present work we continue our lattice field theory investigation of the finite-$\Nf$ GN model by studying the formation of baryons, their spatial distribution and their relation to the chiral condensate. As a preparatory step we also discuss a linear coupling of lattice fermions to the chemical potential.
\end{abstract}

\maketitle


\section{Introduction}

In recent years experiments provided many interesting insights concerning strongly interacting matter at high density (see e.g.\ Ref.\ \cite{Friman:2011zz} for a comprehensive review). On the theoretical side our present understanding of the QCD phase diagram at non-zero chemical potential $\mu$ is to a large extent based on conjectures relying on physical intuition, on model calculations and on effective low energy descriptions \cite{Halasz:1998qr,Fukushima:2010bq}, while reliable ab-initio results are still missing, mostly due to the infamous sign problem in lattice-QCD. Even though there are a number of interesting approaches, which led to considerable progress to mitigate or solve the sign problem, such 
as using complex Langevin algorithms \cite{parisi_ComplexProbabilities_1983,klauder_CoherentstateLangevinEquations_1984,Damgaard:1987rr,Aarts:2009uq,Aarts:2009uq} or thimble methods \cite{Cristoforetti:2012su,Cristoforetti:2013wha,fujii_HybridMonteCarlo_2013}, 
finding more suitable variables \cite{savit_DualityFieldTheory_1980,deForcrand:2014tha,gattringer_NewDevelopmentsDual_2014} or 
refining the density of states approach \cite{gocksch_SimulatingLatticeQCD_1988,gocksch_QuenchedHadronicScreening_1988,gattringer_ApproachesSignProblem_2016a,langfeld_Densityofstates_2017,gattringer_NewDensityStates_2019},
a better understanding of lattice-QCD at finite baryon density is certainly an urgent problem. Urgent, for example, since our colleagues from gravitational wave astronomy and astrophysics are in need of more reliable equations of state of strongly interacting matter at baryon density $n_B$ up to several times the nuclear density $n_0 \approx 0.17 \, \textrm{fm}^{-3}$.

It has been conjectured that in QCD at low temperature and large baryon density there is an inhomogeneous crystalline phase. This conjecture is based on mean field calculations in various effective four Fermi theories indicating the existence of such an inhomogeneous phase
\cite{schnetz_PhaseDiagramGross_2004,deForcrand:2006zz,Nickel:2009wj,Carignano:2010ac,Buballa:2014tba}.
The underlying mean-field (or Hartree-Fock like) approximation becomes exact in the limit of 
an infinite number of fermion flavors $\Nf$, since in this limit quantum fluctuations are negligible.

Mean field approximations are also common in condensed matter physics. For example, for the GN model considered in the present work, the mean field phase diagram with homogeneous and inhomogeneous phases has been known in the condensed matter community \cite{fulde_SuperconductivityStrongSpinExchange_1964,larkin_NonuniformStateSuperconductors_1965}
long before it has been rediscovered in particle physics \cite{thies_RevisedPhaseDiagram_2003,schnetz_PhaseDiagramGross_2004}. 

More recently, interesting models implementing the breaking of
translational invariance
-- for example by charge density waves, dynamical defects or by 
magnetic fields -- have been proposed and studied within the holographic framework \cite{Ammon:2019wci,Baggioli:2020edn}.

At present it is largely unknown, whether crystalline phases exist in effective four Fermi theories at finite number of fermion flavors, or whether quantum fluctuations lead to a qualitatively different phase structure. In a recent work \cite{Lenz:2020bxk} we performed lattice field theory simulations of the GN model in $1+1$ dimensions with $\Nf = 2,8,16$ and found clear evidence for the existence of an inhomgeneous phase, qualitatively similar to that in the limit $\Nf \rightarrow \infty$. In the present work we continue our investigation of the GN model in $1+1$ dimensions at a finite number of fermion flavors and focus on baryonic excitations at low temperature and large chemical potential. We investigate their spatial distribution as well as their relation to the chiral condensate


\section{Chemical Potential for Lattice Fermions}

The continuum Lagrangian density of the (Euclidean)
GN model with vanishing bare mass is given by
\begin{align}
\mathcal{L}_\psi=
\bar\psi \ii \left(\slashed{\partial} + \mu\gamma^{0}\right)\psi + \frac{g^2}{2\Nf}(\bar{\psi}\psi)^2 \, ,
\label{eq:contLagr}
\end{align}
where $\mu$ denotes the chemical potential for the 
conserved baryon number. To be able to perform
the fermion integration one follows Hubbard and 
Stratonovich by introducing a fluctuating auxiliary
scalar field $\sigma$ to linearize the operator
$\bar{\psi}\psi$ in the interaction term 
\begin{align}
\mathcal{L}_\sigma&=
\bar\psi\ii D\psi+\frac{\Nf}{2 g^2} \sigma^2 \, , \quad D=\left( \slashed{\partial}+\sigma+\mu\gamma^{0} \right) \, .
\label{eq:lattLagr}
\end{align}
The four Fermi term in Eq.\ (\ref{eq:contLagr}) is recovered
after eliminating $\sigma$ by its equation of motion or
equivalently by integrating over $\sigma$ in the functional
integral. Translation invariance of
$\D\sigma_x$ in the (well-defined) functional
integral $\prod \D \sigma_x$ for the lattice model implies the Ward identity
\begin{align}
\label{eq:ward1}
\frac{\Nf}{g^{2}} \left\langle \sigma_x \right\rangle = \Big\langle \left(\bar{\psi}\psi\right)_x \Big\rangle \, .
\end{align}

Keeping as many global symmetries of the continuum model as possible
in a discretization can be crucial to obtain a lattice model with the correct continuum limit. Thus we shall discretize the operator 
$D$ using the chiral and doubler-free SLAC
derivative \cite{drell_StrongcouplingFieldTheories_1976,Bergner:2007pu}.
While non-local SLAC fermions must not be used
to discretize a field theory with local gauge symmetries \cite{,karsten_AxialSymmetryLattice_1978,karsten_VacuumPolarizationSLAC_1979,karsten_LatticeFermionsSpecies_1981}, they have
been used successfully in various scalar-field theories
and fermionic theories with global symmetries only \cite{Cohen:1983nr,Wozar:2011gu,Flore:2012xj,
wellegehausen_CriticalFlavourNumber_2017,Lang:2018csk,lenz_AbsenceChiralSymmetry_2019}. 
In addition to using SLAC fermions to simulate the GN model at finite $\Nf$, we have cross-checked our results with a discretization based on naive fermions. This
fermion species is chiral as well but describes
$2^{d}$ doublers in $d$ dimensions. 
More details can be found in Ref.\ \cite{Lenz:2020bxk}.

Besides our preceding paper \cite{Lenz:2020bxk}, we are not aware of any work, in which SLAC fermions have been
used to study fermion systems at finite density. Thus, we begin with comparing the thermodynamics of a gas of free 
massive fermions in a spatial box of size $L$ in the continuum and 
on the lattice with different fermion discretiztations.
A straightforward calculation yields the grand partition function 
at inverse temperature $\beta$ and chemical potential $\mu$
in the continuum
\begin{gather}
\hspace{-1pt}\ln Z_{c} = \sum_{k}\left( \beta E_{k}+\ln \left( 1+e^{-\beta \left( E_{k}-\mu \right)} \right)+\mu\to -\mu \right)
\end{gather}
with single-particle energies $E_{k}^{2}=k^{2}+m^{2}$ 
depending on the spatial wave number
$k = 2 \pi n / L$, $n \in \Z$
and mass $m$. The corresponding baryon density is
\begin{align}
\begin{split}
 n_{B,c} & = \frac{\mathrm{d}\ln Z_c}{\mathrm{d}\mu} \\
& = \frac{1}{L}\sum_{k}\left( \frac{1}{1+e^{\beta\left( E_{k}-\mu \right)}}-\frac{1}{1+e^{\beta\left( E_{k}+\mu \right)}} \right) \, .
\end{split}
\end{align}
Note that the sum over all Matsubara frequencies 
has already been performed, i.e.\ the continuum limit in (imaginary) time
direction is already implied, while truncating the sum over $k$ is conceptually similar to a finite lattice spacing in spatial direction. Since the SLAC
derivative discretizes the continuum dispersion relation up to the maximal 
momentum given by the inverse lattice spacing, the finite (truncated) 
sum is also the result for free, massive SLAC fermions discretized 
in spatial direction only.

This is not what is implemented in lattice Monte Carlo simulations where 
also the (imaginary) time is discretized. Asymptotically, the error
in truncating the Matsubara sum after $\Nt$ terms (with Matsubara
frequencies symmetrically about the origin) is
$\sim \beta \left( E_{k}\pm \mu \right) / \pi^2\Nt$. Letting
the lattice constant in time direction
$\beta/\Nt\to 0$ at \emph{fixed} $k$, we can neglect this error. However, 
$E_{k}$ will eventually become large in the sum over $k$ and higher order 
corrections will contribute, if the temporal ``cutoff'' $\Nt/\beta$ is 
not sent to infinity \emph{before} taking the limit $L/\Ns\to 0$. 
In the particular case of a uniform continuum limit $\Nt=\Ns\to \infty$, 
we pick up the following correction terms in $1+1$ dimensions:
\begin{align}
& \lim_{N_s=N_t\to\infty}\ln Z = \ln Z_{c} - \frac{\mu^2}{4\pi}\\
\label{EQN531} & \lim_{N_s=N_t\to\infty}n_{B} = n_{B,c}-\frac{\mu}{2\pi} \, .
\end{align}
As argued above, the expressions on the left
correspond to the continuum limit of free
massive SLAC fermions,\footnote{When letting $\Nt=\Ns\to\infty$ at fixed box
size. A detailed calculation of the correction term at finite lattice
spacing, possibly different in temporal and spatial direction, can be found in appendix~\ref{APP567}.}
for which the chemical potential enters the Lagrangian linearly 
via $\ii \mu\bar{\psi}\gamma^0\psi$
as it does in the continuum, see e.g.\ Eq.\ (\ref{eq:contLagr}).
We conclude that introducing the chemical potential linearly 
as in the continuum theory yields the correct partition function 
up to a ($\mu$-dependent) constant and can thus be used in Monte Carlo simulations. In fact, since
SLAC fermions couple fermion fields at 
well-separated lattice sites it would be difficult to introduce an
exponentially coupled $\mu$ for all hopping terms in
the Lagrangian. In appendix~\ref{APP567} we show Eq.\ (\ref{EQN531}), i.e.\ that the baryon
density computed from a lattice action with 
linearly coupled $\mu$ has to be corrected 
by the constant $+\mu / 2 \pi$.
This is actually not a defect of SLAC fermions but
just expresses the fact that conventionally one first performs
the continuum limit in time direction and afterwards in
the spatial directions to arrive at the well-known expression
for the thermodynamic potentials at finite temperature and density.
\begin{figure}
\includegraphics[width=\linewidth]{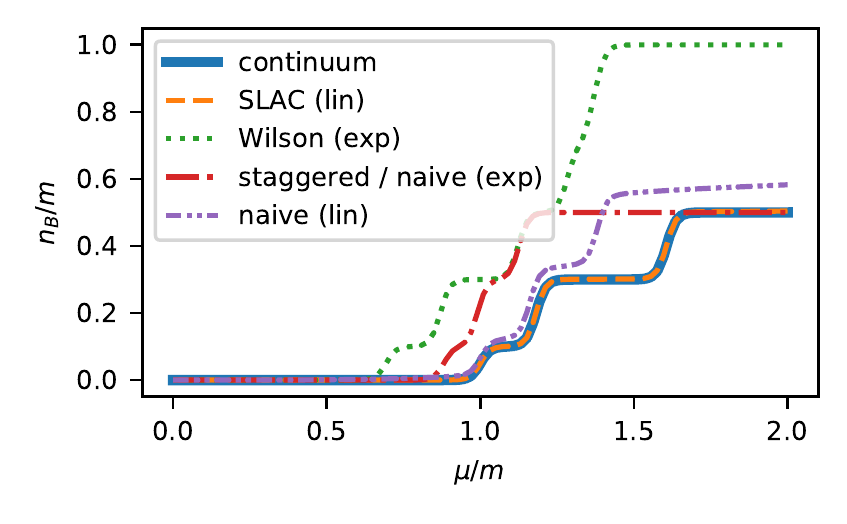}
\caption{\label{f:nBfree}Baryon density $n_B$ as a function of $\mu$ for free massive fermions discretized on an $\Nt \times \Ns=64\times 10$ lattice 
of extent $L m=10$, $\beta m=64$. Note that the baryon density $n_B$ is normalized by multiplication with a 
factor $1/\Nf$, where $\Nf$ also includes possible doublers. For SLAC fermions the additive linear coupling according to Eq.\ (\ref{EQN531}) has been used,  
for Wilson fermions and for staggered fermions only exponential coupling and for naive fermions we distinguish between conventional exponential coupling (exp) and additive linear coupling (lin). Note that staggered fermions and naive fermions with exponential coupling lead to identical results.}

\end{figure}
A comparison of free massive baryon densities for various commonly used lattice discretizations is depicted in \fref{f:nBfree}.  The (properly corrected) SLAC result is almost indistinguishable from the continuum result. In passing, we note the following: 
\begin{itemize}
\item A similar analysis for naive fermions reveals that 
introducing a chemical potential as additive linear term
requires the same correction as for SLAC fermions 
(when $\beta/\Nt$ and $L/\Ns$ approach zero simultaneously)
and thus could  also be used.
\item For $d>2$ space-time dimensions the known
correction term diverges in the continuum limit. 
In $4$ spacetime dimensions it diverges quadratically
\cite{Hasenfratz:1983ba}.
A practical
renormalization scheme on the lattice could then be to determine the 
constant $c$ via
\begin{align}
0=\frac{\mathrm{d}\left( n_{B}-2c\mu \right)}{\mathrm{d}\mu}\bigg|_{\mu=0}
\end{align}
($c$ is finite at finite lattice spacing $a$) and subtracting $c\mu^{2}$ from the partition function.
It has been demonstrated in Ref.\ \cite{Gavai:2014lia} that this
method of divergence removal works in (quenched) QCD.
This is different from earlier attempts to eliminate the divergencies
by suitably modifying the
lattice action
\cite{Bilic:1983fc,hasenfratz_CHEMICALPOTENTIALLATTICE_1983,gavai_ChemicalPotentialLattice_1985}.

\item 
The standard exponential coupling of the chemical potential for naive
fermions has significantly larger discretization errors than the
linear coupling (if corrected properly). In particular, a linearly 
coupled chemical potential yields quite accurately the position of the 
first step, i.e.\ the fermion mass. 

\item In appendix~\ref{APP567} we explicitly calculate
the correction term for $\Nt\approx\Ns\gg 1$ for non-interacting
fermions. We have observed that this correction
is insensitive to the interaction and hence it is sufficient 
to subtract the same term in the GN model \cite{Lenz:2020bxk}. It 
seems that this statement holds true in higher dimensions as well, where
the correction terms are UV-divergent. This has been observed in
numerical simulations \cite{Gavai:2014lia}, but a general 
proof of this interesting observation in QCD and interacting GN models appears to be still missing.

\end{itemize}
We emphasize once more, that the $\mu$-dependent correction terms
 are not lattice artifacts -- they may also appear in continuum theories, depending on how divergent integrals are treated (cf.\ the detailed discussion below Eq.\ (\ref{EQN531})).
We have crosschecked results obtained within this project using naive fermions with a conventional exponentially coupled chemical potential (see Ref.\ \cite{Lenz:2020bxk}) as this is the established method of introducing
a chemical potential in lattice field theory.
Since the exponentially coupled $\mu$ couples to the exactly
conserved charge on the lattice, no correction terms are needed \cite{Bilic:1983fc}.


\section{\label{SEC003}Baryonic matter}

\begin{figure}
\includegraphics[width=\linewidth]{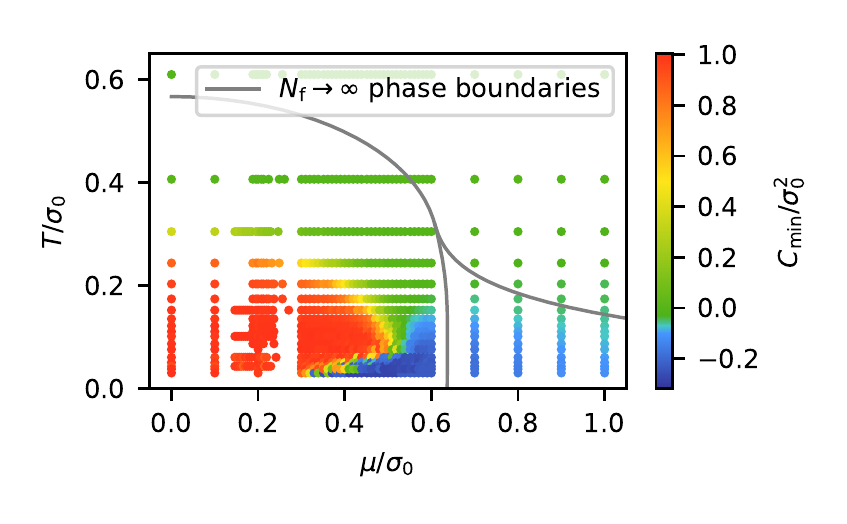}
\caption{\label{f:pd}Phase diagram of the $1+1$-dimensional GN model for $\Nf = 8$ (SLAC fermions, $a \approx 0.410 / \sigma_0$, $\Ns = 63$; figure taken from Ref.\ \cite{Lenz:2020bxk}). The homogeneously broken phase, the symmetric phase and the inhomogeneous phase are colored in red, green and blue, respectively. For comparison also the $\Nf \rightarrow \infty$ phase boundaries are shown as gray lines.}
\end{figure}

In a recent paper \cite{Lenz:2020bxk} we studied the phase structure of the GN model in $1+1$ dimensions for $\Nf = 8$ using lattice field theory and SLAC fermions with exact chiral symmetry. The resulting phase diagram is shown in \fref{f:pd} in units of 
\begin{align}
\label{EQN696} \sigma_0 = \lim_{L \rightarrow \infty} \langle |\overline{\sigma}| \rangle\Big|_{\mu=0,T=0}
\end{align}
with
\begin{align}
\label{EQN697} \overline{\sigma} = \frac{1}{\Nt \Ns} \sum_{t,x} \sigma(t,x) \, .
\end{align}
As expected, we identified a homogeneously broken phase with non-zero constant chiral condensate at small chemical potential and low temperature and a symmetric phase at high temperature.\footnote{Note that our numerical results did not allow to decide whether these regions in the $\mu$-$T$ plane are phases in a strict thermodynamical sense or rather regimes, which strongly resemble phases. In any case, throughout this paper we denote these regions as ``phases''.} Most interestingly, however, we also found a phase, where the spatial correlator
\begin{align}
\label{C-corr} C(x) = \big\langle c(x) \big\rangle \, , \quad
c(x) = \frac{1}{\Nt \Ns} \sum_{t,y} \sigma(t,y+x) \sigma(t,y)
\end{align}
is an oscillating function (see e.g.\ \fref{f:nBsigma2num}, bottom). As a simple observable to distinguish the three phases we used
\begin{align}
C_{\min} = \min_x \, C(x) \, ,
\end{align}
where
\begin{align}
C_{\min} \begin{cases}
\gg 0 & \text{inside the homogeneously broken phase} \\
\approx 0 & \text{inside the symmetric phase} \\
< 0 & \text{inside the inhomogeneous phase}
\end{cases}
\end{align}
(see \fref{f:pd}).

\begin{figure}
\includegraphics[width=\linewidth]{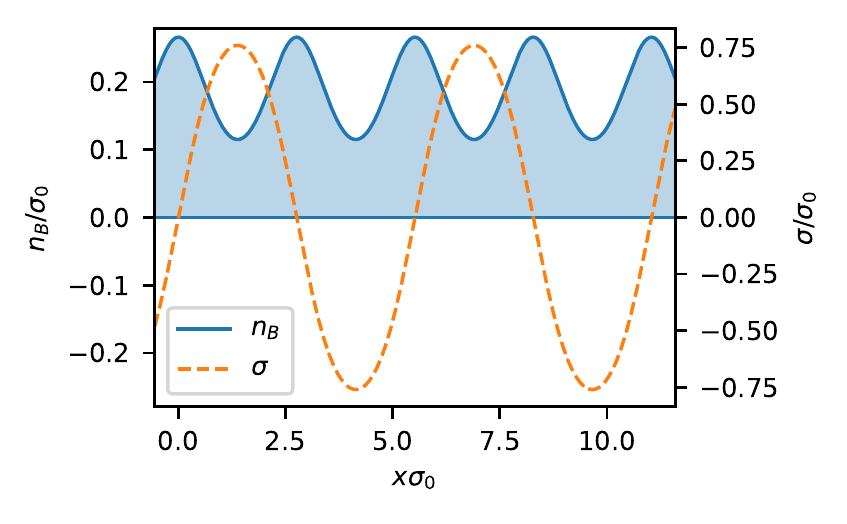}
\caption{\label{f:crystal}$\Nf \rightarrow \infty$ results from Ref.\ \cite{thies_RelativisticQuantumFields_2006a} for the condensate $\sigma(x)$ and the baryon density $n_B(x)$ at $(\mu/\sigma_0 , T/\sigma_0) = (0.700 , 0)$, i.e.\ inside the inhomogeneous phase (see Ref.\ \cite{thies_RelativisticQuantumFields_2006a}, Eqs.\ (54) and (80)).}
\end{figure}

The GN model can be solved analytically in the semi-classical approximation or, equivalently, in the limit \mbox{$\Nf \rightarrow \infty$} (see e.g.\ Refs.\ \cite{Dashen:1975xh,thies_RevisedPhaseDiagram_2003,schnetz_PhaseDiagramGross_2004,thies_RelativisticQuantumFields_2006a}) and it is known that extrema of the effective action
\begin{align}
\label{EQN601} S_\textrm{eff} = \frac{1}{2 g^2} \int\! \mathrm{d}^2x \; \sigma^2 - \ln \det D \, ,
\end{align}
which one obtains by using $\mathcal{L}_\sigma$ from Eq.\ (\ref{eq:lattLagr}) and integrating over the fermions in the partition function, are not only given by $\sigma = \textrm{const}$. For example in Ref.\ \cite{schnetz_PhaseDiagramGross_2004} it was shown that at large chemical potential and small temperature a spatially oscillating function $\sigma(x)$ minimizes the free energy. For each cycle of the oscillation $\Nf$ fermions or antifermions, which can be interpreted as baryons, are located in the region of minimal $\sigma^2$,
i.e.\ where the sign of $\sigma$ changes. This implies breaking of translational symmetry and a crystal of baryons (as shown in \fref{f:crystal}).

In the present work we investigate, whether traces of such a baryonic crystal are also present in the GN model with a finite number of fermion flavors. For all plots shown in the following we performed computations with $\Nf = 8$ flavors of SLAC fermions, lattice spacing $a \approx 0.410 / \sigma_0$ and $\Ns = 63$ lattice sites in spatial direction, corresponding to a periodic spatial direction of extent $L = \Ns a \approx 25.8 / \sigma_0$. We use the same lattice setup and rational
HMC algorithm as in our preceding paper and
refer for technical details to Ref.~\cite{Lenz:2020bxk} and to Appendix C in
Ref.~\cite{Lenz:2019qwu}, where the same setup was used. Not addressed
in these references is the issue of possibly existing exceptional configurations with zero
modes of the Dirac operator, which cannot be
ruled out. We note, however, that we are not considering a gauge theory, where such zero modes are protected by topology.
Indeed we did not encounter any problems in our simulations, which indicate the presence of exceptional configurations in our ensembles.
Note that there is also no sign problem, even for $\mu \neq 0$, because the determinant of the Dirac operator $D$
in Eq. (\ref{eq:lattLagr}) for $\Nf=8$ is always real and non-negative (see Ref.\ \cite{Lenz:2020bxk} for details). 

From the extensive set of simulations we carried out in Ref.\ \cite{Lenz:2020bxk} for different $a$ and $L$, we expect that both lattice discretization errors and finite volume corrections are small. In particular we observed that the size and shape of the inhomogeneous phase is stable, even when varying the lattice spacing by a factor of $\approx 2$ and the spatial volume by a factor of $\approx 4$ (see Fig.\ 8 in Ref.\ \cite{Lenz:2020bxk}). This clearly indicates that the inhomogeneous phase is not an artifact of either the finite lattice spacing or the finite spatial volume. Note that in Ref.\ \cite{Lenz:2020bxk} we also performed computations with $\Nf = 8$ flavors of naive fermions, to check and to confirm our numerical results.


\subsection{\label{SEC567}Correlation of the baryon density and the condensate}

We start by investigating the location of the fermions relative to the spatially oscillating condensate inside the inhomogeneous phase. It is important to note that the effective action \eqref{EQN601} is invariant under spatial translations. Therefore, field configurations, which are spatially shifted relative to each other, i.e.\ $\sigma(t,x)$ and $\sigma(t,x+\delta x)$, contribute with the same weight $e^{-S_\textrm{eff}}$ to the partition function and, thus, should be generated with the same probability by the HMC algorithm. Consequently, simple observables like $\langle \sigma(x) \rangle$ or $\langle n_B(x) \rangle$, where
\begin{align}
\label{EQN602} n_B = \frac{\ii \bar{\psi} \gamma^0 \psi}{\Nf} \, ,
\end{align}
are not suited to detect an inhomogeneous condensate or baryon density in a lattice simulation, because destructive interference should lead to $\langle \sigma(x) \rangle = 0$ and $\langle n_B(x) \rangle = \textrm{const}$, even in cases, where all field configurations exhibit spatial oscillations with the same wavelength. An observable, which does not suffer from destructive interference and is able to exhibit information about possibly present inhomogeneous structures, is the spatial correlation function of $\sigma(x)$, as defined in Eq.\ (\ref{C-corr}) (for a more detailed discussion see section~4.3 of Ref.\ \cite{Lenz:2020bxk}). Similarly, the spatial correlation function of the baryon density and the squared condensate,
\begin{align}
\label{Cp-corr} C_{n_B \sigma^2}(x) = \bigg\langle \frac{1}{\Nt \Ns} \sum_{t,y} n_B(t,y+x)\,\sigma^2(t,y) \bigg\rangle \, ,
\end{align}
\begin{figure*}
	\begin{subfigure}[t]{\columnwidth}
		\includegraphics[width=\linewidth]{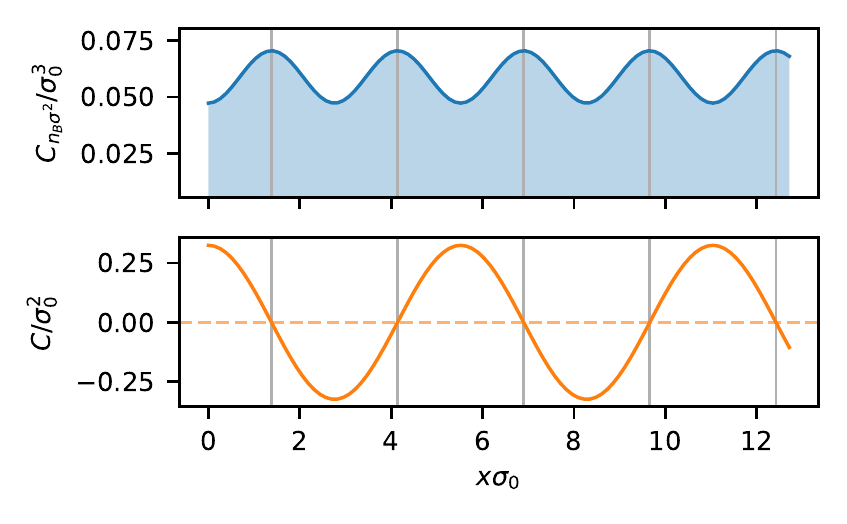}
		\caption{\label{f:nBsigma2anal}Analytical results for $\Nf\to\infty$ and $(\mu/\sigma_0 , T/\sigma_0) = (0.700 , 0.038)$ according to Refs.\ \cite{schnetz_PhaseDiagramGross_2004,Thies:2019}.}
	\end{subfigure}
	\begin{subfigure}[t]{\columnwidth}
		\includegraphics[width=\linewidth]{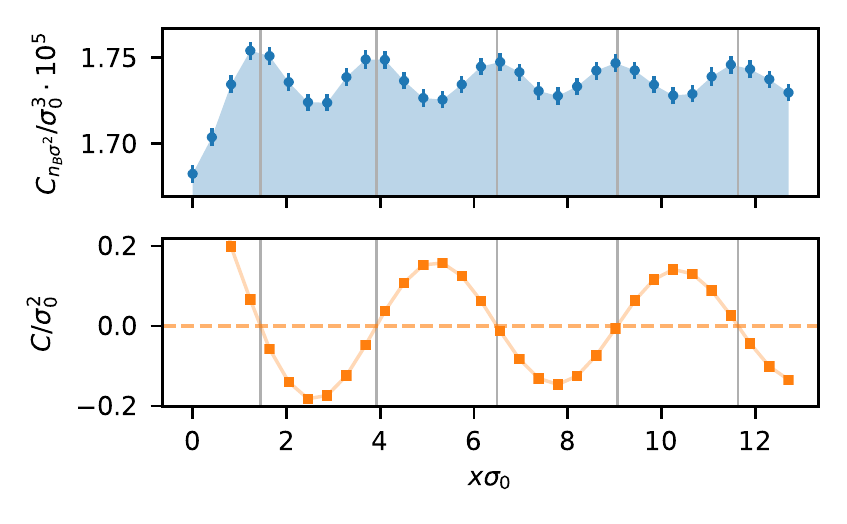}
		\caption{\label{f:nBsigma2num}Lattice field theory results for $\Nf=8$ and $(\mu/\sigma_0 , T/\sigma_0) = (0.700 , 0.038)$.}
	\end{subfigure}
	\caption{\label{f:nBsigma2}The spatial correlation functions $C_{n_B \sigma^2}(x)$ (top) and $C(x)$ (bottom) as defined in Eqs.\ \eqref{Cp-corr} and \eqref{C-corr}. The vertical gray lines indicate the roots of $C(x)$.}
\end{figure*}
can provide insights on the location of the fermions relative to the extrema of the condensate inside an inhomogeneous phase. For $\Nf \rightarrow \infty$ the maxima of $\sigma^2(x)$ coincide with the minima of $n_B(x)$ and vice versa, as one can read off from \fref{f:crystal}. Thus, for $\Nf \rightarrow \infty$ the correlator $C_{n_B \sigma^2}(x)$ has minima at $n \lambda / 2$ and maxima at $(n +1/2) \lambda / 2$, where $\lambda$ is the wavelength of both $\sigma^2(x)$ and $n_B(x)$ (see \fref{f:nBsigma2anal}, where $C_{n_B \sigma^2}(x)$ and $C(x)$ are shown for $(\mu/\sigma_0 , T/\sigma_0) = (0.700 , 0.038)$). Our corresponding lattice results for $\Nf = 8$ at the same chemical potential and temperature exhibit an almost identical behavior (see \fref{f:nBsigma2num}). We interpret this as clear signal that baryons are centered at the roots of the condensate $\sigma$, where $\sigma$ is a periodically oscillating function (we have investigated the latter in detail in our preceding work \cite{Lenz:2020bxk}). Thus separations between neighboring baryons should all be similar, which is reminiscent to the baryonic crystal found at $\Nf \rightarrow \infty$.


\subsection{Baryon number and its relation to the condensate}

In this section we study the baryon number
\begin{align}
\label{EQN_B} B = \left\langle \int\! \mathrm{d}x \; n_B \right\rangle
\end{align}
with the baryon density $n_B$ as defined in Eq.\ \eqref{EQN602} and investigate its relation to the average number of cycles of the oscillating condensate $\sigma$. Inside the finite periodic lattice with extent $L$ we have defined and computed
\begin{align}
\label{EQN_nu} \nu_\textrm{max} = \frac{L \langle |k_\textrm{max}| \rangle}{2 \pi} \, .
\end{align}
By $k_\textrm{max}$ we denote the dominant momentum of $c(x)$ (see Eq.\ \eqref{C-corr}), i.e.\ that $k$, which maximizes the absolute value of the Fourier transform $\tilde{c}(k)$.

\begin{figure}
\includegraphics[width=\columnwidth]{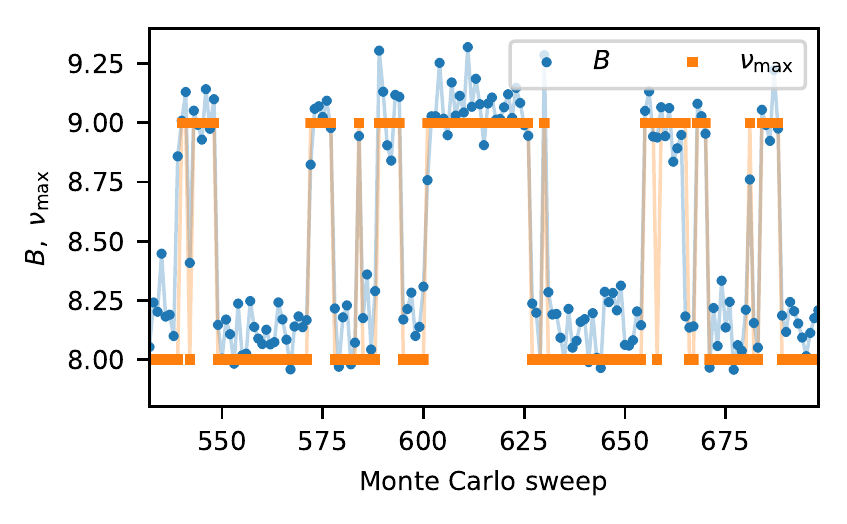}
\caption{\label{FIG_T0.038_2}Monte Carlo histories of $B$ and of $\nu_\textrm{max}$ at $(\mu/\sigma_0 , T/\sigma_0) = (1.10,0.038)$ after thermalization.}
\end{figure}

The central result of this subsection is that $B$ and $\nu_\textrm{max}$ are almost identical on each field configuration, i.e.\@ even when omitting the average $\langle \ldots \rangle$ over all generated field configurations in the definitions \eqref{EQN_B} and \eqref{EQN_nu}. In particular at small $T$ there is almost perfect agreement. This is illustrated in \fref{FIG_T0.038_2}, where we show Monte Carlo histories of $B$ and of $\nu_\textrm{max}$ at $(\mu/\sigma_0 , T/\sigma_0) = (1.10,0.038)$ after thermalization. We interpret this as strong indication that the GN model with $\Nf = 8$ behaves very similar to the GN model in the limit $\Nf \rightarrow \infty$, where $B = \nu_\textrm{max}$.

In the following we elaborate on this further by presenting and discussing results obtained at two different values for the temperature, $T/\sigma_0 \approx 0.076$ and $T/\sigma_0 \approx 0.038$.


\subsubsection{\label{SEC578}Temperature $T/\sigma_0 \approx 0.076$}

\begin{figure}
\includegraphics[width=\columnwidth]{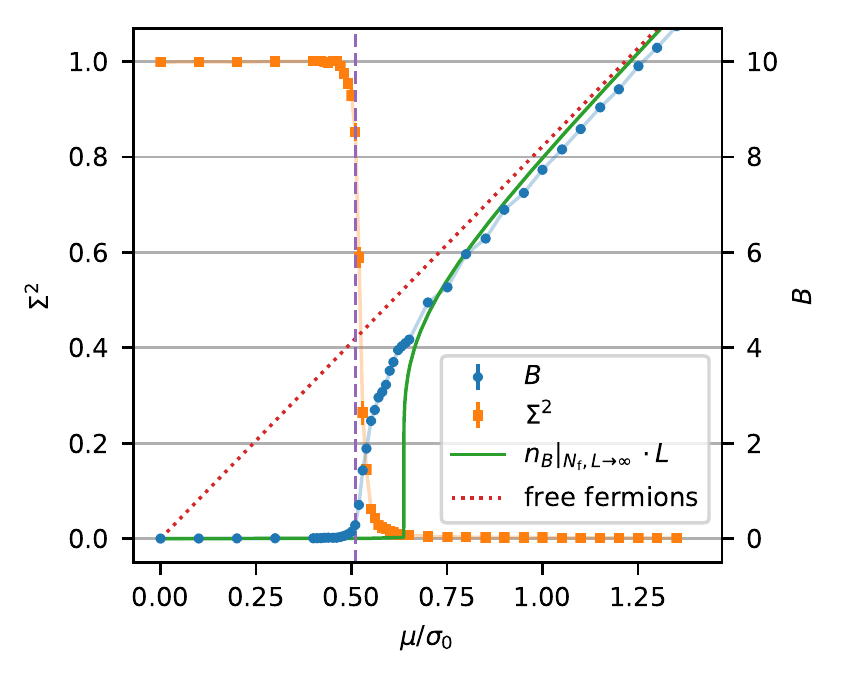}
\caption{\label{FIG_T0.076_1}
Baryon number $B$ and $\Sigma^2$ as functions of the chemical potential $\mu$ at temperature $T/\sigma_0 \approx 0.076$. The green curve represents $n_B|_{\Nf,L \to \infty} L$. The dashed vertical line indicates the phase transition at $\mu/\sigma_0 \approx 0.51$.}
\end{figure}

For temperature $T/\sigma_0 \approx 0.076$ we show both $B$ and
\begin{align}
\Sigma^2 = \frac{\langle \overline{\sigma}^2 \rangle}{\sigma_0^2}
\end{align}
with $\sigma_0$ and $\overline{\sigma}$ as defined in Eq.\ (\ref{EQN696}) and Eq.\ (\ref{EQN697}) as functions of the chemical potential $\mu$ in \fref{FIG_T0.076_1}. At \mbox{$\mu/\sigma_0 \approx 0.51$} both quantities indicate in a consistent way the phase transition between the homogeneously broken and the inhomogeneous phase. While $B \approx 0$ for small $\mu$, $B$ suddenly starts to increase at $\mu/\sigma_0 \approx 0.51$. At roughly the same $\mu$ value $\Sigma^2$ rapidly drops from around $1$ to $0$. Note that $B$ is quite similar to $n_B|_{\Nf,L \to \infty} L$ (the green solid line in \fref{FIG_T0.076_1}), where $n_B|_{\Nf,L \to \infty}$ is the analytical infinite volume result for the $\Nf \to \infty$ baryon density according to Ref.\ \cite{schnetz_PhaseDiagramGross_2004}. However, the phase transition for $\Nf = 8$ takes place at smaller $\mu/\sigma_0 \approx 0.51$ compared to $\mu/\sigma_0 \approx 2 / \pi \approx 0.64$ for $\Nf \rightarrow \infty$, where (at $T = 0$) a baryon corresponds to a kink-antikink field configuration $\sigma$ with energy $2 \sigma_0 / \pi$ per fermion (see e.g.\ Refs.\ \cite{Dashen:1975xh,karsch_GrossNeveuModelFinite_1987}). This can also be seen in the phase diagram shown in \fref{f:pd} and implies that the baryon mass (per fermion flavor) is somewhat smaller compared to $\Nf\to\infty$. Note that at finite $\Nf$ a smaller homogeneously broken phase is expected, because of fluctuations in $\sigma$, which increase disorder (see the detailed discussion in our preceding work \cite{Lenz:2020bxk}).

The careful reader might already note the remnants of the stair-like low temperature behavior discussed later on in section~\ref{SEC590}.
Moreover, at large $\mu$ the $\Nf = 8$ result and the $\Nf \rightarrow \infty$ result for the baryon number $B$ are very similar, where the latter approaches the corresponding result for free fermions, as noted in Ref.\ \cite{thies_RevisedPhaseDiagram_2003}.

\begin{figure}
\includegraphics[width=\columnwidth]{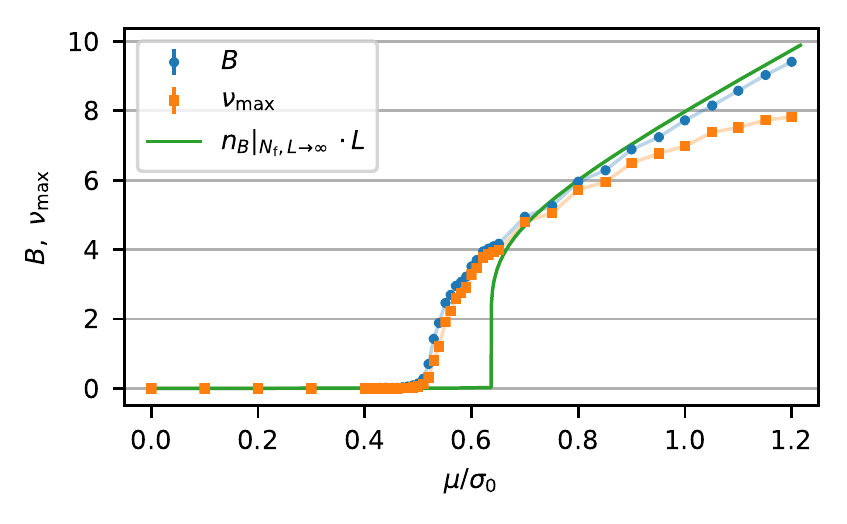}
\caption{\label{FIG_T0.076_2}$B$ and $\nu_\textrm{max}$ as functions of the chemical potential $\mu$ at temperature $T/\sigma_0 \approx 0.076$. The green curve represents $n_B|_{\Nf,L \to \infty} L$.}
\end{figure}

\begin{figure}
\includegraphics[width=\columnwidth]{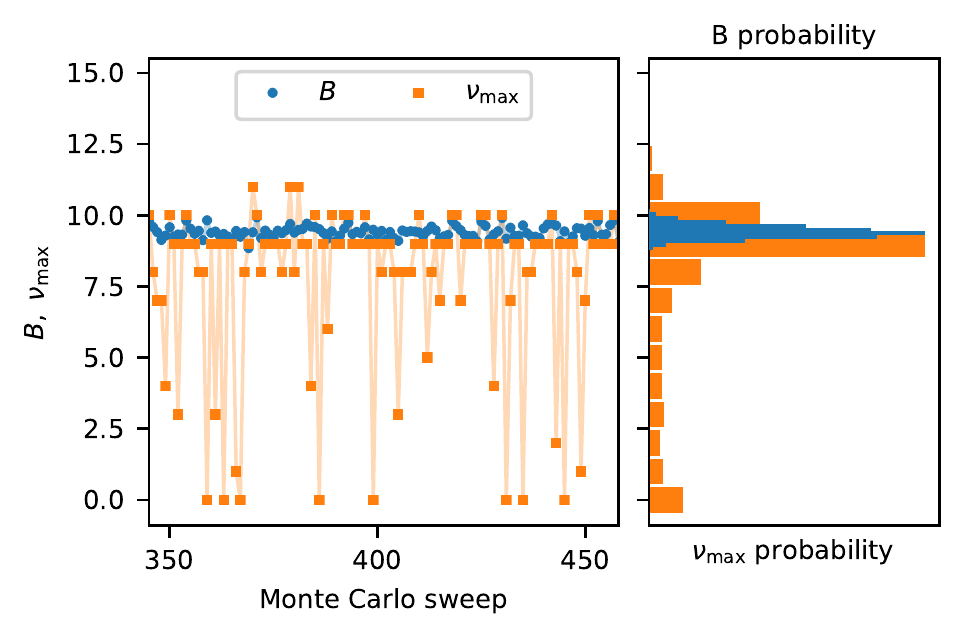}
\caption{\label{FIG_T0.076_3}Left plot: Monte Carlo histories of $B$ and of $\nu_\textrm{max}$ at $(\mu/\sigma_0 , T/\sigma_0) = (1.20,0.076)$ after thermalization. Right plot: Histograms of $B$ and of $\nu_\textrm{max}$ approximating their probability distribution.}
\end{figure}

In \fref{FIG_T0.076_2} we show $B$ and $\nu_\textrm{max}$ as functions of the chemical potential $\mu$. The two curves are quite similar, which is a strong indication that the finite-$\Nf$ theory is qualitatively well-described by the semi-classical $\Nf \rightarrow \infty$ picture.
Note that $\nu_\textrm{max}$ is slightly below $B$ for $\mu/\sigma_0 \gtapprox 0.51$. The reason can be seen in \fref{FIG_T0.076_3}, where thermalized Monte Carlo histories of $B$ and $\nu_\textrm{max}$ are shown for $\mu/\sigma_0 = 1.20$ (analogous plots for other values of $\mu/\sigma_0 \gtapprox 0.51$ are similar). On the majority of generated field configurations $B$ and $\nu_\textrm{max}$ agree rather well. However, $B$ is an extremely stable quantity, while $\nu_\textrm{max}$ exhibits sizable fluctuations on around $25 \%$ of the generated field configurations, mostly fluctuations towards small values, significantly below the median of $\nu_\textrm{max}$. Such fluctuations are more common for larger values of $\mu$. This is expected from the Fourier transformed spatial correlation function, which we investigated in Ref.\ \cite{Lenz:2020bxk} in detail,
and also reflected by the corresponding histogram for $\nu_\textrm{max}$ shown in the right part of \fref{FIG_T0.076_3}. To summarize, \fref{FIG_T0.076_2} indicates that the baryon number $B$ is quite similar to the number of cycles of the spatial oscillation of the condensate $\sigma$, as for $\Nf \rightarrow \infty$. This observation supports our conclusions above that, also at finite $\Nf$, baryons are the relevant excitations of the GN model and that their number is closely related to the shape of the condensate $\sigma$.

\subsubsection{\label{SEC590}Temperature $T/\sigma_0 \approx 0.038$}

\begin{figure}
\includegraphics[width=\columnwidth]{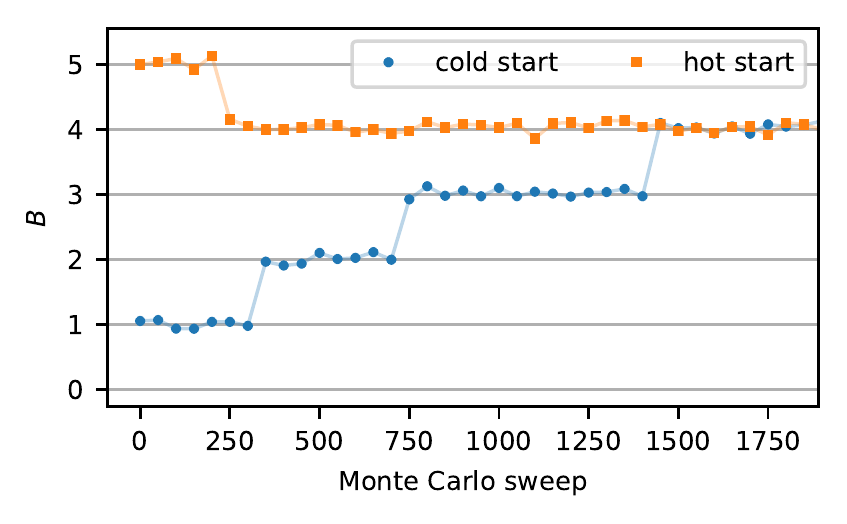}
\caption{\label{FIG_T0.038_1}Monte Carlo histories of $B$ at $(\mu/\sigma_0 , T/\sigma_0) = (0.65,0.038)$ for a cold start and a hot start. For the sake of readability, only every 50$^{\text{th}}$ measurement is shown.}
\end{figure}

Autocorrelatios at $T/\sigma_0 \approx 0.038$ turned out to be rather large in our simulations, in particular near the boundary of the homogeneously broken phase and the inhomogeneous phase, in the region $0.40 \ltapprox \mu / \sigma_0 \ltapprox 0.65$. This is illustrated in \fref{FIG_T0.038_1}, where we compare Monte Carlo histories of $B$ at $\mu/\sigma_0 = 0.65$ for a cold start (each field variable $\sigma(t,x) = 1$) and a hot start (each field variable drawn randomly from a Gaussian distribution with mean 0). After around $1500$ Monte Carlo sweeps the two Monte Carlo histories eventually converge and the simulations seem to have thermalized. Nevertheless the autocorrelation time is quite large, of the order of the average number of Monte Carlo sweeps needed to create or annihilate a baryon, i.e.\ $\gtapprox 500$. This is sizable compared to the typical number of Monte Carlo sweeps, between $2000$ and $10000$, we are able to carry out for each simulation with our available HPC resources.
Thus, for $\mu/\sigma_0 \approx 0.65$ the errors we show for our results might be somewhat underestimated.

\begin{figure}
figures/\includegraphics[width=\columnwidth]{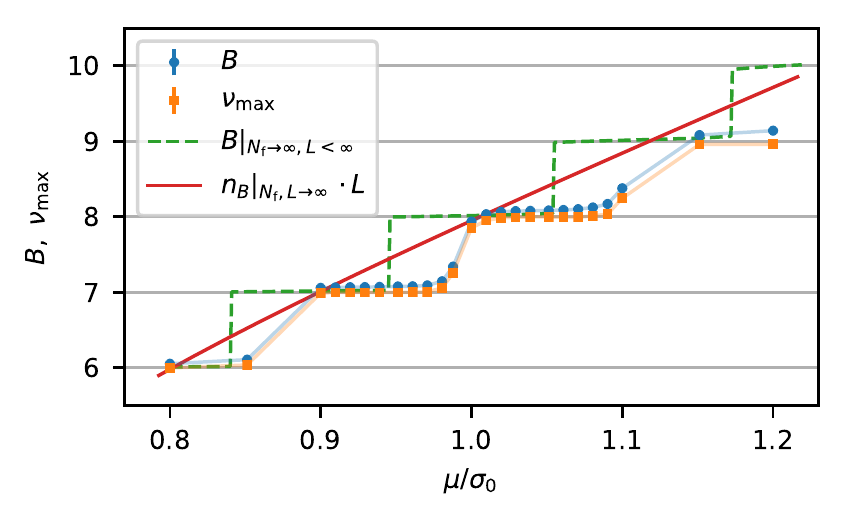}
\caption{\label{FIG_T0.038_6}$B$ and $\nu_\textrm{max}$ as functions of the chemical potential $\mu$ at temperature $T/\sigma_0 \approx 0.038$.
The green curve represents the corresponding $\Nf \rightarrow \infty$ lattice field theory result for $B$ at spatial lattice extent $L \approx 25.8 / \sigma_0$ (the same extent used in our simulations), while the purple curve 
is the analytically known $\Nf \rightarrow \infty$ continuum result for $n_B|_{\Nf,L \to \infty} L$.}
\end{figure}

For larger $\mu/\sigma_0$, i.e.\ farther away from the phase boundary, autocorrelation times become smaller. For example, in \fref{FIG_T0.038_2} we show the Monte Carlo histories of $B$ and of $\nu_\textrm{max}$ at $\mu/\sigma_0 = 1.10$ after thermalization. For almost all field configurations $B \approx \nu_\textrm{max}$ and their values are either close to $8$ or close to $9$. Even though we only show 150 Monte Carlo sweeps, there are many transtions between $B \approx \nu_\textrm{max} = 8$ and $B \approx \nu_\textrm{max} = 9$, indicating that the HMC algorithm is able to frequently increase or decrease the number of cycles of the spatial oscillation of the condensate $\sigma$.
In \fref{FIG_T0.038_6} we show $B$ and $\nu_\textrm{max}$ as functions of $\mu$ in the range $0.80 \leq \mu/\sigma_0 \leq 1.20$. Both quantities exhibit a very clear stair-like behavior with the steps corresponding to integers.
The steps of $\nu_\textrm{max}$ 
(which are more pronounced at low temperature due to
less thermal fluctuations) are explained by
the semi-classical commensurability constraint, 
namely that the wave length of the 
periodic condensate must divide the box length
$L$. In the limit $\Nf\to\infty$ the baryon number is equal to the cycles
of the condensate. Thus, a second baryon appears at chemical potential
$\mu_2 > \mu_c \approx (2 / \pi) \sigma_0$, a third baryon at
$\mu_3>\mu_2$, etc. Hence, with increasing $\mu$ the mean separations
of baryons decreases which leads to more interaction.
That the semi-classical picture explains the simulations so 
well is a further 
indication that the GN model at $\Nf = 8$ is quite similar 
to the GN model for $\Nf \rightarrow \infty$, i.e.\ quantum fluctuations at $\Nf = 8$ seem to be rather weak.

We note that the stair-like behavior is a consequence of the finite spatial extent $L$. To support this we minimized the SLAC regularized GN action with a specific ansatz for the chiral condensate,
\begin{align}
\sigma(x) = A \cos\left(\frac{2\pi}{L} q x \right) \, , \quad q\in\mathbb{N} \, ,
\end{align}
in the variables $A$ and $q$. This ansatz is a reasonable 
approximation for the considered values of $\mu$ as the analytically known 
chiral condensate in the $\Nf,L \to \infty$ case rapidly reduces to a cos-shape 
for increasing chemical potential. This enables us to calculate $B$ for
$\Nf \to \infty$ and finite $L$. The result for $a \approx 0.410 / \sigma_0$ and $\Ns = 63$ corresponding to $L = \Ns a \approx 25.8 / \sigma_0$ (i.e.\ the same lattice spacing and extent used in our simulations) exhibits clear steps as shown in \fref{FIG_T0.038_6}. 
The steps disappear for $L \rightarrow \infty$ as shown 
by $n_B|_{\Nf,L \to \infty} L$, which is also plotted in \fref{FIG_T0.038_6}.

\begin{figure}
	\includegraphics[width=\columnwidth]{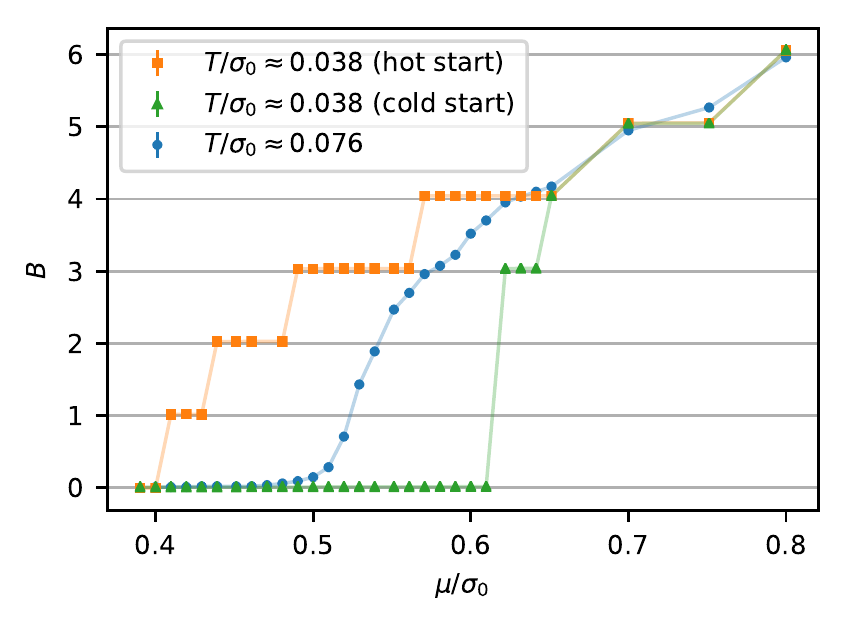}
	\caption{\label{FIG_T0.038_3}$B$ as a function of $\mu$ for $T/\sigma_0 \approx 0.038$ obtained from cold starts and from hot starts. For comparison we also show the corresponding result at $T/\sigma_0 \approx 0.076$, where thermalization and large autocorrelations do not cause problems.}
\end{figure}

For $\mu / \sigma_0 \ltapprox 0.65$ autocorrelation times are extremely large and we were not able to reach thermal equilibrium in our simulations. This is shown in \fref{FIG_T0.038_3}, where we present results for $B$ obtained from cold starts and from hot starts. For $0.4 \ltapprox \mu/\sigma_0 \ltapprox 0.65$ the cold and the hot curves differ and the largest discrepancy is observed close to the phase transition around $\mu/\sigma_0 \approx 0.51$. We expect the true result for $B(\mu)$ to be somewhere between the two curves, i.e.\ the cold and the hot results represent lower and upper bounds for $B(\mu)$. 
This is also supported by 
our simulation results for $B(\mu)$ at $T/\sigma_0 \approx 0.076$ (see section~\ref{SEC578}), which is bounded by the cold and the hot curves obtained at $T/\sigma_0 \approx 0.038$ in this range.
\fref{FIG_T0.038_3} as well as the exceedingly long
autocorrelation times remind us of hysteresis effects 
near a first order transition. From the good agreement with 
the semi-classical picture, one could conjecture that in 
a finite volume the probability distribution $e^{-S_{\mathrm{eff}}}/Z$ 
has two peaks (due to the commensurability constraint) 
which leads to the observed hysteresis effects. 
The problem will probably go away for very large volumes, 
but for high-precision simulations on finite lattices
near the transition improved algorithms are needed, which
support the creation and annihilation of extended baryons.

We remark that consequences of the large autocorrelations are also visible in the phase diagram shown in \fref{f:pd}. In the problematic region, i.e.\ for $0.40 \ltapprox \mu/\sigma_0 \ltapprox 0.65$ and $T/\sigma_0 \ltapprox 0.05$, the boundary between the homogeneously broken phase and the inhomogeneous phase suddenly turns towards the origin, which amounts to an inhomogeneous phase larger than expected and qualitatively different from the $\Nf \rightarrow \infty$ boundary. Knowing that all simulations for this phase diagram were started with hot field configurations, this behavior can now be understood as a thermalization problem (see \fref{FIG_T0.038_3}, where the result for $B(\mu)$ obtained with a hot start incorrectly indicates the phase boundary at a rather small value for $\mu$. 

\section{Conclusions}

In this work we investigated the distribution of the baryon density
$n_B$ in the $\Nf=8$ GN model enclosed in a finite box of
size $L$ at finite chemical potential $\mu$.
The simulations were performed with chiral SLAC
fermions. We compared with recent results on
the spatial inhomogeneities in the $1+1$-dimensional GN model 
\cite{Lenz:2020bxk}, which is interpreted as modulation of baryonic matter 
density as observed in the $\Nf\to\infty$ limit of the model
\cite{schnetz_PhaseDiagramGross_2004} and which is well-known 
in solid state physics \cite{fulde_SuperconductivityStrongSpinExchange_1964,larkin_NonuniformStateSuperconductors_1965}. 
Since translation symmetry is inherent to
Monte Carlo simulations on finite lattices (the phase of a quasi-periodic
configuration is a collective parameter)
we could not measure the baryon density  $n_B(x)$
directly. Instead we found a strong correlation between the dominant 
wave number of the spatial inhomogeneities and the baryon number. 
This is clear evidence for a region in the phase diagram
corresponding to a regime of modulated baryonic matter. Via this detour 
we explicitly circumvented the question about the breaking
of translation symmetry. The delicate question whether we found a rigid 
baryon crystal (as seen in the large-$\Nf$ limit) or rather a baryonic 
liquid \cite{chandler_satisticalmechanics_1987},
where the baryons have a preferred separation locally, but are 
disordered on large scales, has been 
addressed in Ref.\
\cite{Lenz:2020bxk} and needs further investigations
with improved algorithms.

Our results shed further light on what can happen in quantum 
field theories at large fermion densities. In particular, 
it shows that mean-field and large-$\Nf$ approximations may contain
more (hidden) information on the physics at finite $\Nf$ than one
would expect. This is reassuring, 
since in particle physics and even more so in solid state physics 
we often rely on these approximations. Our results may also be 
of relevance in condensed matter systems, 
e.g.\ for large, almost 1-dimensional polymers
\cite{Chodos:1993mf}.

However, it is still unclear if the above results are relevant
for QCD. On the one hand, we established that the interpretation 
as baryonic matter is not spoiled by taking quantum fluctuations into 
account. On the other hand, although recent numerical lattice-studies 
of the GN model in $2+1$ dimensions and for $\Nf\to\infty$
spotted inhomogenous condensates, the spatial modulation
is related to the cutoff scale and disappears
in the continuum limit \cite{Winstel:2019zfn,Narayanan:2020uqt}. 
Clearly, if this happens in the limit $\Nf\to\infty$, then we cannot
expect a breaking of translation invariance for a finite 
number of flavours.
Thus, extending our numerical studies and simulations
to fermion systems in higher dimensions is an important task. 
Interesting candidates are for example the Nambu-Jona-Lasinio model or the quark-meson model in $3+1$ dimensions.


\acknowledgements

We acknowledge useful discussions with Martin Ammon, Michael Buballa, Philippe de Forcrand, Holger Gies, Felix Karbstein, Adrian K\"onigstein, Maria Paola Lombardo, Dirk Rischke, Alessandro Sciarra, Lorenz von Smekal, Michael Thies and Marc Winstel on various aspects of fermion theories and spacetime symmetries.

We thank Philippe de Forcrand for bringing reference \cite{chandler_satisticalmechanics_1987} to our attention.

We also thank Marc Winstel for providing the code base which was used to obtain the finite Volume $\Nf\to\infty$ lattice results.

Furthermore we thank Michael Thies for providing us the analytic expression for the spatial correlator in the $\Nf,L\to\infty$ limit.

J.J.L.\ and A.W.\ have been supported by the Deutsche Forschungsgemeinschaft (DFG) under Grant No.\ 406116891 within the Research Training Group RTG 2522/1. L.P.\ and M.W.\ acknowledge support by the Deutsche Forschungsgemeinschaft (DFG, German Research Foundation) through the CRC-TR 211 ``Strong-interaction matter under extreme conditions'' -- project number 315477589 -- TRR 211. M.W.\ acknowledges support by the Heisenberg Programme of the Deutsche Forschungsgemeinschaft (DFG, German Research Foundation) -- project number 399217702.

Calculations on the GOETHE-HLR high-performance computers of the Frankfurt University as well as on the ARA cluster of the University of Jena supported in part by DFG grants INST 275/334-1 FUGG and INST 275/363-1 FUGG were conducted for this research. We would like to thank HPC-Hessen, funded by the State Ministry of Higher Education, Research and the Arts, and Andr\'e Sternbeck from the Universit\"atsrechenzentrum Jena for programming advice.


\appendix

\section{\label{APP567}Continuum limit for free fermions with	linearly coupled chemical potential $\mu$}

In this appendix we calculate the correction term, which must be added to the baryon density, when using a linearly coupled chemical potential and removing the lattice cutoff as it is typically done in lattice field theory.

We consider non-interacting fermions linearly coupled to $\mu$ enclosed in a $(d-1)$-dimensional spatial box of linear extent $L$ subject to periodic boundary conditions. The allowed wave numbers of the corresponding fermion field at finite temperature $T=1/\beta$ are
\begin{equation}
k \in \{ (k_0,\vk) \} = \bigg\{ \bigg(\omega_{n},\frac{2\pi}{L} \vn\bigg)  \bigg\} \, , \quad (n,\vn) \in \Z^d \label{mu1}
\end{equation}
with Matsubara frequencies
\begin{equation}
\omega_n = \frac{2 \pi}{\beta} \bigg(n + \frac{1}{2}\bigg) \, .
\end{equation}
The corresponding eigenvalues of the free Dirac operator with chemical potential are
\begin{equation}
\lambda^\pm_k = m \pm \sqrt{(\mu + \ii k_0)^2 - \vk^2} \, ,
\end{equation}
such that
\begin{equation}
\lambda_k^+ \lambda_k^- = (k_0 - \ii\mu)^2 + E_{\vk}^2 \, , \quad E_{\vk}^2 = \vk^2 + m^2 \, . \label{mu3}
\end{equation}
Because of spin, the degeneracy of the eigenvalues is $\mathcal{C}=2^{[d/2]-1}$, where $[x]$ denotes the greatest integer less than or equal to $x$.

The logarithm of the grand partition function $Z(\mu)$ divided by $\beta V$ is the pressure $p$. The $\mu$-derivative of $p$ is the baryon density,
\begin{align}
n_B = \frac{\partial p}{\partial\mu} = \frac{\partial}{\partial\mu} \frac{\ln Z(\mu)}{\beta V} = \frac{\mathcal{C}}{\beta V} \sum_{\vk} v(E_{\vk}) \, , \label{nb1}
\end{align}
with
\begin{align}
v(E_\vk) = \frac{\partial}{\partial \mu} \sum_{\omega_n} \ln \lambda_k^+ \lambda_k^- \, . \label{nb2}
\end{align}
The sum defining $v(E_\vk)$ is convergent for any fixed $\vk$. But when we sum over the spatial momenta, $E_\vk$ (which is the positive square root
of $E_\vk^2$)
becomes arbitrarily large and we will show that removing the cutoffs in frequency space and momentum space does not commute.

To this end, we regularize the sum over the Matsubara frequencies (\ref{nb2}) by only admitting $\Nt$ frequencies. It is convenient to choose these frequencies symmetric to the origin (which is only
possible for even $\Nt$), i.e.\@ we restrict $\omega_n$ according to
\begin{equation}
\vert\omega_n\vert \leq
\frac{2\pi N'_\text{t}}{\beta} \, , \quad N'_\text{t} = \frac{\Nt-1}{2} \, .
\end{equation}
To calculate the truncated series over the $\omega_n$, denoted by $v(E_\vk,\Nt)$, we combine terms with $\pm \omega_n$ and obtain
\begin{equation}
v(E_\vk,\Nt) = \sum_{\vert n+\frac{1}{2} \vert \leq N'_\text{t}} \frac{\mu-E_\vk}{\omega_n^2+(\mu-E_\vk)^2} + \big(E_\vk \to -E_\vk\big) \, . \label{vsum}
\end{equation}
where $(E_\vk\to- E_\vk)$ represents the previous term with opposite
sign of the energy.
$v(E_\vk,\Nt)$ is finite for $\Nt\to\infty$ and can be calculated,
\begin{equation}
v(E_\vk) = v(E_\vk,\infty) =\frac{\beta}{e^{\beta(E_{\vk}-\mu)}+1} - \big(\mu \to -\mu\big) \, . \label{vseries}
\end{equation}
Inserting (\ref{vseries}) into (\ref{nb1}) yields the baryon density
\begin{equation}
n_B = \frac{\mathcal{C}}{V} \sum_{\vk} \bigg(\frac{1}{1+e^{\beta(E_\vk-\mu)}} - \frac{1}{1+e^{\beta(E_\vk+\mu)}}\bigg) \, . \label{eq:Zmu}
\end{equation}
An integration with respect to the chemical potential gives the pressure of the Fermi gas. The integration constant is the divergent contribution of the quantum fluctuations at zero temperature and zero chemical potential,
\begin{align}
\frac{pV}{\mathcal{C}} = \sum_{\vk} \bigg(E_\vk + \frac{1}{\beta} \ln\big(1+e^{-\beta(E_\vk-\mu)}\big) + (\mu \to -\mu)\Big) \, .
\end{align}
To summarize: if we first perform the continuum limit in Euclidean time
direction, which means sum over all $n \in \Z$ in Eq.\ (\ref{nb2}), then 
we get the sum $\sum E_\vk$  at $\mu = T =0$ plus the \emph{finite sum}
known from quantum statistics.

Since the sum over the Matsubara frequencies is only conditionally convergent and the sum over the $\vk$ is divergent, we get a different result, when we remove the cutoff in the Matsubara frequencies together with the cutoff in the spatial momenta, as one does in lattice field theory. To show that, we consider the difference between the series (\ref{vseries}) and the finite sum (\ref{vsum}),
\begin{align}
\Delta v(E_\vk,\Nt) & = v(E_\vk,\infty) - v(E_\vk,\Nt) \nonumber \\
\begin{split}
& = \frac{\beta}{\pi} F_{\frac{N_{t}+1}{2}}\bigg(\frac{\beta(\mu-E_\vk)}{2\pi}\bigg)
\\
&\quad - \frac{\beta}{\pi}F_{\frac{N_{t}+1}{2}}\bigg(\frac{\beta(\mu+E_\vk)}{2\pi}\bigg) \, , \label{diffF}
\end{split}
\end{align}
where we introduced the function
\begin{align}
F_{\kappa}(z) = \sum_{n=0}^\infty \frac{z}{z^2 + (n+\kappa)^2} = \sum_{n,m=0}^\infty \frac{(-1)^{m}z^{2m+1}}{(n+\kappa)^{2m+2}}
\end{align}
with $\kappa = (N_t+1)/2$. For large $\kappa$ the sum over $n$ is approximately given by
\begin{equation}
\label{eq:kappaExpansion}
\sum_{n=0}^\infty \frac{1}{(n+\kappa)^{s+1}} = \frac{1}{s\kappa^s}-\frac{1}{2 \kappa^{s+1}}
+ \dots
\end{equation}
In the following we focus on free fermions in $1+1$ dimensions, where only the first term gives a finite contribution to the error. The other terms are suppressed by inverse powers of $\Nt$. Thus, keeping 
the relevant term we arrive at
\begin{equation}
F_{\kappa}(z) = \sum_{m=0}^\infty \frac{(-1)^{m}}{2m+1} \bigg(\frac{z}{\kappa}\bigg)^{2m+1} \, . \label{eq:F}
\end{equation}
For large spatial momenta we have $\mu \ll E_\vk$ and
\begin{equation}
(\mu+E_\vk)^{2m+1} + (\mu-E_\vk)^{2m+1} \sim 2(2m+1) \mu E_\vk^{2m} \, ,
\end{equation}
such that
\begin{equation}
\Delta v(E_\vk,\Nt) = \frac{2\beta \mu}{\pi} \sum_{m=0}^\infty (-1)^m \bigg(\frac{\beta}{2\pi\kappa}\bigg)^{2m+1}E_\vk^{2m} \, . \label{dvsum1}
\end{equation}
To study the error for the baryon density
\begin{equation}
\Delta n_B = \frac{1}{\beta L} \sum_{\vk} \Delta v(E_\vk,\Nt) \label{dvsum2}
\end{equation}
(in 2 spacetime dimensions $\mathcal{C}=1$) as a function of the temporal and spatial cutoffs, we cut off the spatial momenta as
\begin{equation}
\vert k_1 \vert \leq
\frac{2\pi \Ns'}{L} \, , \quad \Ns'=\frac{\Ns-1}{2} \, .
\end{equation}
For convenience we choose the spatial momenta symmetric to the origin. Inserting Eq.\ (\ref{dvsum1}) into the regularized sum (\ref{dvsum2}) we can calculate the leading term with the help of
\begin{equation}
\sum_{n=-\Ns'}^{\Ns'} E_\vk^{2m} \sim \frac{2}{2m+1} \bigg(\frac{2\pi}{L}\bigg)^{2m} \bigg(\frac{N_s}{2}\bigg)^{2m+1} \, . \label{err17}
\end{equation}
This way we end up with
\begin{align}
\Delta n_B = \frac{2\mu}{\pi^2}\sum_{m=0}^\infty \frac{(-1)^m}{2m+1} \bigg(\frac{\Ns\beta}{\Nt L}\bigg)^{2m+1} = \frac{2\mu}{\pi^2} \text{atan} \bigg(\frac{\Ns \beta}{\Nt L}\bigg) .
\end{align}
If we regularize the system on a lattice with the same lattice constant $a$ in temporal and spatial direction then the argument of $\text{atan}$ is equal to $1$ and we obtain
\begin{equation}
\Delta n_B = \frac{\mu}{2\pi} \, .
\end{equation}
To summarize: If we use a linearly coupled chemical potential for free fermions on the lattice, then we must subtract from the resulting baryon density $n_B$ a term linear in $\mu$, in order to recover the result obtained in a conventional continuum calculation. In spacetime dimensions $d>2$ the correction term actually is UV-divergent, since an analogous estimate reveals that the suppression by inverse powers of $\kappa \propto \Nt$ does not anymore balance the sum over the spatial momenta.

Finally, we note that higher order terms in this correction (vanishing for $N_{t}\to\infty$) are now straightforward to compute. For future reference, we just show the $\mathcal{O}(a)$ correction on a hypercubic lattice with lattice spacing $a$ ($\beta = a\Nt,\ L=a\Ns$):
\begin{equation}
\Delta n_{B} = \frac{\mu}{2\pi}\left( 1-\frac{4a
}{\pi \beta} \right) + \mathcal{O}\left( a^{2} \right).
\end{equation}
In higher orders in $a$, also terms $\sim \mu^{2n+1},\ n\in\N,$ appear.

\bibliography{baryonLetter}

\begin{thebibliography}{10}
\providecommand{\url}[1]{\texttt{#1}}
\providecommand{\urlprefix}{URL }
\input{babelbst.tex}
\newcommand{\Capitalize}[1]{\uppercase{#1}}
\newcommand{\capitalize}[1]{\expandafter\Capitalize#1}
\providecommand{\eprint}[2][]{\url{#2}}

\bibitem{Lenz:2020bxk}
{Lenz, Julian and Pannullo, Laurin and Wagner, Marc and Wellegehausen, Bj\"orn
  and Wipf, Andreas}, \emph{{Inhomogeneous phases in the Gross-Neveu model in
  1+1 dimensions at finite number of flavors}}, Phys. Rev. D \textbf{101}
  (2020), \bblno{}~9 094512, \eprint{2004.00295}.

\bibitem{Friman:2011zz}
B.~Friman, C.~H\"ohne, J.~Knoll, S.~Leupold, J.~Randrup, R.~Rapp \bbland{}
  P.~Senger, \emph{The CBM physics book. Compressed baryonic matter in
  laboratory experiments}, \bblvol{} 814, Springer-Verlag Berlin Heidelberg
  (2011).

\bibitem{Halasz:1998qr}
A.~M. Halasz, A.~D. Jackson, R.~E. Shrock, M.~A. Stephanov \bbland{} J.~J.~M.
  Verbaarschot, \emph{{On the phase diagram of QCD}}, Phys. Rev. D \textbf{58}
  (1998) 096007, \eprint{hep-ph/9804290}.

\bibitem{Fukushima:2010bq}
K.~Fukushima \bbland{} T.~Hatsuda, \emph{{The phase diagram of dense QCD}},
  Rept. Prog. Phys. \textbf{74} (2011) 014001, \eprint{1005.4814}.

\bibitem{parisi_ComplexProbabilities_1983}
G.~Parisi, \emph{On complex probabilities}, Phys. Lett. B \textbf{131} (1983)
  393.

\bibitem{klauder_CoherentstateLangevinEquations_1984}
J.~R. Klauder, \emph{{Coherent State Langevin Equations for Canonical Quantum
  Systems With Applications to the Quantized Hall Effect}}, Phys. Rev. A
  \textbf{29} (1984) 2036.

\bibitem{Damgaard:1987rr}
P.~H. Damgaard \bbland{} H.~Huffel, \emph{{Stochastic Quantization}}, Phys.
  Rept. \textbf{152} (1987) 227.

\bibitem{Aarts:2009uq}
G.~Aarts, E.~Seiler \bbland{} I.-O. Stamatescu, \emph{{The Complex Langevin
  method: When can it be trusted?}}, Phys. Rev. D \textbf{81} (2010) 054508,
  \eprint{0912.3360}.

\bibitem{Cristoforetti:2012su}
M.~Cristoforetti, F.~Di~Renzo \bbland{} L.~Scorzato (AuroraScience), \emph{{New
  approach to the sign problem in quantum field theories: High density QCD on a
  Lefschetz thimble}}, Phys. Rev. D \textbf{86} (2012) 074506,
  \eprint{1205.3996}.

\bibitem{Cristoforetti:2013wha}
M.~Cristoforetti, F.~Di~Renzo, A.~Mukherjee \bbland{} L.~Scorzato, \emph{{Monte
  Carlo simulations on the Lefschetz thimble: Taming the sign problem}}, Phys.
  Rev. D \textbf{88} (2013), \bblno{}~5 051501, \eprint{1303.7204}.

\bibitem{fujii_HybridMonteCarlo_2013}
H.~Fujii, D.~Honda, M.~Kato, Y.~Kikukawa, S.~Komatsu \bbland{} T.~Sano,
  \emph{{Hybrid Monte Carlo on Lefschetz thimbles - A study of the residual
  sign problem}}, JHEP \textbf{10} (2013) 147, \eprint{1309.4371}.

\bibitem{savit_DualityFieldTheory_1980}
R.~Savit, \emph{{Duality in Field Theory and Statistical Systems}}, Rev. Mod.
  Phys. \textbf{52} (1980) 453.

\bibitem{deForcrand:2014tha}
P.~de~Forcrand, J.~Langelage, O.~Philipsen \bbland{} W.~Unger, \emph{{Lattice
  QCD Phase Diagram In and Away from the Strong Coupling Limit}}, Phys. Rev.
  Lett. \textbf{113} (2014), \bblno{}~15 152002, \eprint{1406.4397}.

\bibitem{gattringer_NewDevelopmentsDual_2014}
C.~Gattringer, \emph{{New developments for dual methods in lattice field theory
  at non-zero density}}, PoS \textbf{LATTICE2013} (2014) 002,
  \eprint{1401.7788}.

\bibitem{gocksch_SimulatingLatticeQCD_1988}
A.~Gocksch, \emph{Simulating {{Lattice QCD}} at {{Finite Density}}}, Phys. Rev.
  Lett. \textbf{61} (1988) 2054.

\bibitem{gocksch_QuenchedHadronicScreening_1988}
A.~Gocksch, P.~Rossi \bbland{} U.~M. Heller, \emph{Quenched Hadronic Screening
  Lengths at High Temperature}, Phys. Lett. B \textbf{205} (1988) 334.

\bibitem{gattringer_ApproachesSignProblem_2016a}
C.~Gattringer \bbland{} K.~Langfeld, \emph{{Approaches to the sign problem in
  lattice field theory}}, Int. J. Mod. Phys. A \textbf{31} (2016), \bblno{}~22
  1643007, \eprint{1603.09517}.

\bibitem{langfeld_Densityofstates_2017}
K.~Langfeld, \emph{{Density-of-states}}, PoS \textbf{LATTICE2016} (2017) 010,
  \eprint{1610.09856}.

\bibitem{gattringer_NewDensityStates_2019}
{Gattringer, Christof and Mandl, Michael and T\"orek, Pascal}, \emph{{New
  density of states approaches to finite density lattice QCD}}, Phys. Rev. D
  \textbf{100} (2019), \bblno{}~11 114517, \eprint{1911.05320}.

\bibitem{schnetz_PhaseDiagramGross_2004}
O.~Schnetz, M.~Thies \bbland{} K.~Urlichs, \emph{{Phase diagram of the
  Gross-Neveu model: Exact results and condensed matter precursors}}, Annals
  Phys. \textbf{314} (2004) 425, \eprint{hep-th/0402014}.

\bibitem{deForcrand:2006zz}
P.~de~Forcrand \bbland{} U.~Wenger, \emph{{New baryon matter in the lattice
  Gross-Neveu model}}, PoS \textbf{LAT2006} (2006) 152,
  \eprint{hep-lat/0610117}.

\bibitem{Nickel:2009wj}
D.~Nickel, \emph{Inhomogeneous Phases in the {{Nambu}}-{{Jona}}-{{Lasino}} and
  Quark-Meson Model}, Phys. Rev. D \textbf{80} (2009) 074025,
  \eprint{0906.5295}.

\bibitem{Carignano:2010ac}
S.~Carignano, D.~Nickel \bbland{} M.~Buballa, \emph{{Influence of vector
  interaction and Polyakov loop dynamics on inhomogeneous chiral symmetry
  breaking phases}}, Phys. Rev. D \textbf{82} (2010) 054009,
  \eprint{1007.1397}.

\bibitem{Buballa:2014tba}
M.~Buballa \bbland{} S.~Carignano, \emph{Inhomogeneous Chiral Condensates},
  Prog. Part. Nucl. Phys. \textbf{81} (2015) 39, \eprint{1406.1367}.

\bibitem{fulde_SuperconductivityStrongSpinExchange_1964}
P.~Fulde \bbland{} R.~A. Ferrell, \emph{{Superconductivity in a Strong
  Spin-Exchange Field}}, Phys. Rev. \textbf{135} (1964) A550.

\bibitem{larkin_NonuniformStateSuperconductors_1965}
A.~I. Larkin \bbland{} Y.~N. Ovchinnikov, \emph{Nonuniform State of
  Superconductors}, Sov. Phys. JETP \textbf{20} (1965), \bblno{}~3 762.

\bibitem{thies_RevisedPhaseDiagram_2003}
M.~Thies \bbland{} K.~Urlichs, \emph{{Revised phase diagram of the Gross-Neveu
  model}}, Phys. Rev. D \textbf{67} (2003) 125015, \eprint{hep-th/0302092}.

\bibitem{Ammon:2019wci}
M.~Ammon, M.~Baggioli \bbland{} A.~Jiménez-Alba, \emph{{A Unified Description
  of Translational Symmetry Breaking in Holography}}, JHEP \textbf{09} (2019)
  124, \eprint{1904.05785}.

\bibitem{Baggioli:2020edn}
M.~Baggioli, S.~Grieninger \bbland{} L.~Li, \emph{{Magnetophonons \& type-B
  Goldstones from Hydrodynamics to Holography}}  (2020), \eprint{2005.01725}.

\bibitem{drell_StrongcouplingFieldTheories_1976}
S.~Drell, M.~Weinstein \bbland{} S.~Yankielowicz, \emph{{Strong Coupling Field
  Theories. 2. Fermions and Gauge Fields on a Lattice}}, Phys. Rev. D
  \textbf{14} (1976) 1627.

\bibitem{Bergner:2007pu}
G.~Bergner, T.~Kaestner, S.~Uhlmann \bbland{} A.~Wipf, \emph{{Low-dimensional
  Supersymmetric Lattice Models}}, Annals Phys. \textbf{323} (2008) 946,
  \eprint{0705.2212}.

\bibitem{karsten_AxialSymmetryLattice_1978}
L.~H. Karsten \bbland{} J.~Smit, \emph{{Axial Symmetry in Lattice Theories}},
  Nucl. Phys. B \textbf{144} (1978) 536.

\bibitem{karsten_VacuumPolarizationSLAC_1979}
L.~H. Karsten \bbland{} J.~Smit, \emph{{The Vacuum Polarization With {SLAC}
  Lattice Fermions}}, Phys. Lett. B \textbf{85} (1979) 100.

\bibitem{karsten_LatticeFermionsSpecies_1981}
L.~H. Karsten \bbland{} J.~Smit, \emph{{Lattice Fermions: Species Doubling,
  Chiral Invariance, and the Triangle Anomaly}}, Nuclear Physics B \textbf{183}
  (1981), \bblno{}~1 103 .

\bibitem{Cohen:1983nr}
Y.~Cohen, S.~Elitzur \bbland{} E.~Rabinovici, \emph{A {{Monte Carlo}} Study of
  the {{Gross}}-{{Neveu}} Model}, Nucl. Phys. B \textbf{220} (1983) 102.

\bibitem{Wozar:2011gu}
C.~Wozar \bbland{} A.~Wipf, \emph{{Supersymmetry Breaking in Low Dimensional
  Models}}, Annals Phys. \textbf{327} (2012) 774, \eprint{1107.3324}.

\bibitem{Flore:2012xj}
R.~Flore, D.~Korner, A.~Wipf \bbland{} C.~Wozar, \emph{{Supersymmetric
  Nonlinear O(3) Sigma Model on the Lattice}}, JHEP \textbf{11} (2012) 159,
  \eprint{1207.6947}.

\bibitem{wellegehausen_CriticalFlavourNumber_2017}
{Wellegehausen, Bj\"orn H. and Schmidt, Daniel and Wipf, Andreas},
  \emph{{Critical flavor number of the Thirring model in three dimensions}},
  Phys. Rev. D \textbf{96} (2017), \bblno{}~9 094504, \eprint{1708.01160}.

\bibitem{Lang:2018csk}
T.~C. Lang \bbland{} A.~M. L{\"a}uchli, \emph{{Quantum Monte Carlo Simulation
  of the Chiral Heisenberg Gross-Neveu-Yukawa Phase Transition with a Single
  Dirac Cone}}, Phys. Rev. Lett. \textbf{123} (2019), \bblno{}~13 137602,
  \eprint{1808.01230}.

\bibitem{lenz_AbsenceChiralSymmetry_2019}
{Lenz, Julian and Wellegehausen, Bj\"orn H. and Wipf, Andreas}, \emph{{Absence
  of chiral symmetry breaking in Thirring models in 1+2 dimensions}}, Phys.
  Rev. D \textbf{100} (2019), \bblno{}~5 054501, \eprint{1905.00137}.

\bibitem{Hasenfratz:1983ba}
P.~Hasenfratz \bbland{} F.~Karsch, \emph{{Chemical Potential on the Lattice}},
  Phys. Lett. \textbf{125B} (1983) 308.

\bibitem{Gavai:2014lia}
R.~V. Gavai \bbland{} S.~Sharma, \emph{{Divergences in the quark number
  susceptibility: The origin and a cure}}, Phys. Lett. \textbf{B749} (2015) 8,
  \eprint{1406.0474}.

\bibitem{Bilic:1983fc}
N.~Bilic \bbland{} R.~V. Gavai, \emph{{On the Thermodynamics of an Ideal Fermi
  Gas on the Lattice at Finite Density}}, Z. Phys. C \textbf{23} (1984) 77.

\bibitem{hasenfratz_CHEMICALPOTENTIALLATTICE_1983}
P.~Hasenfratz \bbland{} F.~Karsch, \emph{{Chemical Potential on the Lattice}},
  Phys. Lett. B \textbf{125} (1983) 308.

\bibitem{gavai_ChemicalPotentialLattice_1985}
R.~Gavai, \emph{{Chemical Potential on the Lattice Revisited}}, Phys. Rev. D
  \textbf{32} (1985) 519.

\bibitem{Dashen:1975xh}
R.~F. Dashen, B.~Hasslacher \bbland{} A.~Neveu, \emph{{Semiclassical Bound
  States in an Asymptotically Free Theory}}, Phys. Rev. D \textbf{12} (1975)
  2443.

\bibitem{thies_RelativisticQuantumFields_2006a}
M.~Thies, \emph{{From relativistic quantum fields to condensed matter and back
  again: Updating the Gross-Neveu phase diagram}}, J. Phys. A \textbf{39}
  (2006) 12707, \eprint{hep-th/0601049}.

\bibitem{Lenz:2019qwu}
J.~Lenz, B.~H. Wellegehausen \bbland{} A.~Wipf, \emph{Absence of Chiral
  Symmetry Breaking in {{Thirring}} Models in 1+2 Dimensions} \textbf{D100},
  \bblno{}~5 054501, \eprint{1905.00137}.

\bibitem{Thies:2019}
M.~Thies, unpublished notes  (2019).

\bibitem{karsch_GrossNeveuModelFinite_1987}
F.~Karsch, J.~B. Kogut \bbland{} H.~Wyld, \emph{{The {Gross-Neveu} Model at
  Finite Temperature and Density}}, Nucl. Phys. B \textbf{280} (1987) 289.

\bibitem{chandler_satisticalmechanics_1987}
D.~Chandler, \emph{Introduction to modern statistical mechanics}, New York :
  Oxford University Press (1987).

\bibitem{Chodos:1993mf}
A.~Chodos \bbland{} H.~Minakata, \emph{{The Gross-Neveu model as an effective
  theory for polyacetylene}}, Phys.\ Lett.\ A \textbf{191} (1994) 39.

\bibitem{Winstel:2019zfn}
M.~Winstel, J.~Stoll \bbland{} M.~Wagner, \emph{{Lattice investigation of an
  inhomogeneous phase of the 2+1-dimensional Gross-Neveu model in the limit of
  infinitely many flavors}}  (2019), \eprint{1909.00064}.

\bibitem{Narayanan:2020uqt}
R.~Narayanan, \emph{{Phase diagram of the large $N$ Gross-Neveu model in a
  finite periodic box}}, Phys. Rev. D \textbf{101} (2020), \bblno{}~9 096001,
  \eprint{2001.09200}.

\end{thebibliography}

\cleardoublepage

\end{document}